\documentclass[sigconf,authorversion,nonacm]{acmart}

\AtBeginDocument{%
  \providecommand\BibTeX{{%
    \normalfont B\kern-0.5em{\scshape i\kern-0.25em b}\kern-0.8em\TeX}}}

\usepackage{subfigure}
\usepackage{comment}
\usepackage{booktabs}
\usepackage{xurl}

\usepackage{amsmath}
\usepackage{amssymb,amsfonts}
\usepackage[ruled,vlined,linesnumbered]{algorithm2e}
\usepackage{algorithmic}
\usepackage{mathrsfs}
\usepackage{graphicx}
\usepackage{textcomp}
\usepackage{xcolor}

\author{Shuyuan Zhang}
\affiliation{
  \institution{Shanghai Jiao Tong University}
  \country{Shanghai, China}
}
\email{zhang-shuyuan@sjtu.edu.cn}

\author{Shu Shan}
\affiliation{
  \institution{Shanghai Jiao Tong University}
  \country{Shanghai, China}
}
\email{shan.shu@sjtu.edu.cn}

\author{Shizhen Zhao}
\affiliation{
  \institution{Shanghai Jiao Tong University}
  \country{Shanghai, China}
}
\email{shizhenzhao@sjtu.edu.cn}

\newtheorem{definition}{Definition}


\begin{document}
\fancyhead{}

\title[Reducing Reconfiguration Time in Optical-Electrical DCNs]{Reducing Reconfiguration Time in Hybrid Optical-Electrical Datacenter Networks (Extended Abstract)}


\begin{abstract}
We study how to reduce the reconfiguration time in hybrid optical-electrical Datacenter Networks (DCNs). With a layer of Optical Circuit Switches (OCSes), hybrid optical-electrical DCNs can reconfigure its logical topologies to better match the on-going traffic patterns, but the reconfiguration time may directly affect the benefits of reconfigurability. The reconfiguration time consists of the topology solver running time and the network convergence time after triggering reconfiguration. However, existing topology solvers either incur high algorithmic complexity or fail to minimize the reconfiguration overhead.

In this paper, we propose a novel algorithm that combines the ideas of bipartition and Minimum Cost Flow (MCF) to reduce the overall reconfiguration time. For the first time, we formulate the topology solving problem as a MCF problem with piecewise-linear cost, which strikes a better balance between solver complexity and solution optimality. Our evaluation shows that our algorithm can significantly reduce the network convergence time while consuming less topology solver running time, making its overall performance superior to existing algorithms. Our code and test cases are available at \cite{repository}.
\end{abstract}





\maketitle

\textcolor{red}{\textit{This article is the extended abstract of our paper presented in APNET 2023. The full paper is available at }}\url{https://dl.acm.org/doi/10.1145/3600061.3600071}.

\section{Introduction}

The interest in hybrid optical-electrical Datacenter Networks (DCNs) has been growing as it offers the capability of performing traffic-aware topology designs. As network bandwidth keeps increasing, building Clos networks is becoming cost-prohibitive \cite{osa}. In fact, the traffic in DCNs is highly skewed and time-varying \cite{network_traffic_characteristics}. By adapting the network topology to the traffic patterns, the performance-cost ratio can be improved. This necessitates a reconfigurable topology, which can be achieved by introducing novel optical network components such as Optical Circuit Switches (OCSes).

We study how to reduce the reconfiguration time in hybrid optical-electrical DCNs. When the DCN traffic pattern changes, we may need to compute a new topology for this traffic pattern, and then reconfigure the DCN from its old topology to this new topology. The reconfiguration time consists of the topology solver running time and the network convergence time after triggering reconfiguration. The former one depends on the algorithmic complexity and the latter one depends on the number of links to be changed during the reconfiguration process. Therefore, we need a topology solver with low algorithmic complexity, while being able to reduce the number of reconfigured links. 

Existing algorithms are limited because they either incur high topology solver running time or yield a highly suboptimal solution that requires many link reconfigurations. The topology optimization problem can be formulated as an Integer Linear Programming (ILP) problem which is hard to solve by directly using an ILP solver. To reduce the algorithmic complexity, \cite{minimal_rewiring} takes advantage of the homogeneity of the DCN physical topologies and presents a greedy minimal-rewiring algorithm using Minimum Cost Flow (MCF), but experimental results show that the total number of rewires can be far from optimal. On the other hand, \cite{patent} presents an algorithm that utilizes the idea of bipartition and achieves a lower number of rewires than the MCF-based algorithm, but it is still based on ILP and can be very slow in practice.

In this paper, we propose an algorithm that combines the advantages of MCF and bipartition to reduce the total number of rewires, while ensuring a polynomial running time. The standard form of an MCF problem has a linear cost for each link. We note that the cost function of each link can be generalized to a convex piecewise linear function. This observation allows us to develop a polynomial algorithm with minimum number of rewires for the case where there are two OCSes. We then generalize this algorithm to the $n$-OCS cases using an iterative decomposition approach. Our evaluation shows that our algorithm exhibits very low algorithmic complexity, while achieving a low rewiring ratio at the same time. Our code and test cases are available at \cite{repository}.

\section{Problem}

An OCS possesses many input and output ports that can be interconnected with (electrical) switches. A complete matching between the input and output ports can be configured within the OCS. The physical connections between the OCS and the switches is referred to as the \textbf{physical topology}, while the corresponding equivalent topology in the absence of the OCS is referred to as the \textbf{logical topology}.

Consider a flat topology comprising of $m$ Top-of-Rack (ToR) switches and $n$ OCSes, where the uplinks of the ToR switches are only connected to the OCSes. The physical topology of the network is characterized by two key parameters, denoted as $a\in \mathbb{Z}_{\ge 0}^{m\times n}$ and $b\in \mathbb{Z}_{\ge 0}^{m\times n}$. Here, $a_{jk}$ represents the number of connections \textit{from} the $k$-th OCS \textit{to} the $j$-th switch, and $b_{ik}$ signifies the number of connections \textit{from} the $i$-th switch \textit{to} the $k$-th OCS. The logical topology is characterized by $c\in \mathbb{Z}_{\ge 0}^{m\times m}$. $c_{ij}$ is the number of equivalent connections from the $i$-th switch to the $j$-th switch. For convenience, we define index sets $I=\{1,2,\dots,m\}$, $J=\{1,2,\dots,m\}$ and $K=\{1,2,\dots,n\}$. 

The matching of all OCSes can be represented by $x\in \mathbb{Z}_{\ge 0}^{m\times m\times n}$. $x_{ijk}$ is the number of equivalent connections from the $i$-th switch to the $j$-th switch established by the forwarding of the $k$-th OCS. Given the physical and logical topology, a feasible matching should satisfy the following constraints.

\begin{equation*}\label{constr:a}
\sum_{i\in I}x_{ijk}= a_{jk}, \quad \forall j\in J, k\in K 
\end{equation*}

\begin{equation*}\label{constr:b}
    \sum_{j\in J}x_{ijk}= b_{ik}, \quad \forall i\in I, k\in K 
\end{equation*}

\begin{equation*}\label{constr:c}
    \sum_{k\in K}x_{ijk}=c_{ij}, \quad \forall i\in I, j\in J 
\end{equation*}

We define the set of all feasible matchings as $S(a,b,c)$. We use $u\in S(a^\prime, b^\prime, c^\prime)$ and $x \in S(a, b, c)$ to represent the old and new matching of OCSes, respectively. Since physical topology seldom changes, we assume that $a^\prime = a$ and $b^\prime = b$.

We may use the minimum number of disconnections to reflect the network convergence time \cite{minimal_rewiring}. Therefore, our target is to solve the following optimization problem\footnote{We define $x^+ = \max\{x, 0\}$.}.
\begin{equation*}
\begin{aligned}
\min_x &\sum_{i\in I}\sum_{j \in J}\sum_{k \in K}(u_{ijk} - x_{ijk})^+ \\
\text{s.t.} &\quad x \in S(a, b, c)
\end{aligned}
\end{equation*}

In this paper, we focus on a special case where the physical topology is \textit{proportional}.

\begin{definition}
\label{definition:proportion}
A physical topology is defined to be \textit{proportional}, if there exist $r \in \mathbb{Z}_{> 0}^n$, $a^\prime, b^\prime \in \mathbb{Z}_{> 0}^m$, such that $a_{jk}=r_k a^\prime_j, \forall j \in J, k \in K$ and $b_{ik}=r_k b^\prime_i, \forall i \in I, k \in K$.
\end{definition}

\section{Algorithm Design}

\subsection{An Exact Polynomial-Time Algorithm for a Special Case} \label{section:exact}

When $n=2$, we can rewrite the objective function and other constraints using the constraint $x_{ij2} = c_{ij} - x_{ij1}$ to obtain the following equivalent problem.

\begin{subequations}
\begingroup
\allowdisplaybreaks
\begin{align}
\min_{x_{ij1}} &\sum_{i\in I}\sum_{j \in J}[(u_{ij1} - x_{ij1})^+ + (u_{ij2} - c_{ij} + x_{ij1})^+] \notag \\
\text{s.t.} &\quad \sum_{i\in I}x_{ij1}= a_{j1}, \quad \forall j\in J \label{constr:two 1} \\
&\quad \sum_{j\in J}x_{ij1}= b_{i1}, \quad \forall i\in I \label{constr:two 2}\\
&\quad x_{ij1} \le c_{ij}, \quad \forall i\in I, j \in J \label{constr:two 3} 
\end{align}
\endgroup
\end{subequations}

The problem is thus equivalent to the following MCF problem. There are $m$ supply nodes $\{s_1, s_2,\dots, s_m\}$ and $m$ demand nodes $\{d_1, d_2, \dots, d_m\}$. The supply node $s_i$ has $b_{i1}$ units of supply, and the demand node $d_j$ has $a_{j1}$ units of demand. This setting models the constraints \eqref{constr:two 1} and \eqref{constr:two 2}. For each pair of $(s_i, d_j)$, consider the function
$$
f_{ij}(x) = (u_{ij1} - x)^+ + (u_{ij2} - c_{ij} + x)^+, \quad x \in [0, c_{ij}].
$$
This is a convex piecewise-linear function. Assume that it has $q$ noncontinuous points $\{x_1, x_2, \dots, x_q\}$ and define $x_0 = 0$, $x_{q+1} = c_{ij}$. Assume that on $[x_{p-1}, x_{p}]$ the slope of $f_{ij}(\cdot)$ is $\alpha_{p}$. Then we add $q+1$ arcs from $s_i$ to $d_j$. For the $p$-th arc, the cost is $\alpha_{p}$ and the capacity is $x_p - x_{p-1}$. This models the objective function and constraint \eqref{constr:two 3}.

Integral MCF problem is a special ILP that can be solved in polynomial time, and in practice the solving is usually fast, so when $n=2$, our problem can be solved efficiently.

\subsection{The General Algorithm}

For general cases where $n > 2$, we can merge some OCSes to be a larger OCS so that the physical topology can be seen as if it has only 2 OCSes. The merging is an approximation because it widens the range of reconfiguration. Then we can solve the approximated problem using the algorithm proposed in Section \ref{section:exact}. To obtain a real feasible solution, we need to decompose the solution on each imaginary OCS, which requires solving two subproblems recursively. We prove the correctness of our algorithm when the physical topology is proportional, and we prove that the worst-case time complexity of our algorithm is $O(m^4 n \log m + m^2n \log n) \approx O(m^4 n \log m)$, if we choose even bipartition at each division step and use the cost-scaling algorithm for solving the MCF problem.

\section{Evaluation}

We compare our algorithm with two existing algorithms, which are referred to as the Greedy MCF Algorithm \cite{minimal_rewiring} and the Bipartition Algorithm \cite{patent}. The experiment is carried out on a software simulator. The test cases are generated using real-world traces of two Facebook datacenter clusters \cite{sigmetrics20complexity}. The evaluation shows that:

\begin{itemize}
    \item Our proposed algorithm outperforms the others in terms of speed, and in some instances, it is up to 10 times faster than the second fastest algorithm, the Greedy MCF Algorithm.
    \item The approximation ratio of our algorithm is comparable to that of the Bipartition Algorithm, and in most cases, our algorithm has a better approximation ratio than the Greedy MCF Algorithm.
\end{itemize}

To conclude, our algorithm achieves significant speed improvements while maintaining a comparable approximation ratio.

  \bibliographystyle{ACM-Reference-Format}
  \bibliography{bibliography}


\begin{thebibliography}{6}


\ifx \showCODEN    \undefined \def \showCODEN     #1{\unskip}     \fi
\ifx \showDOI      \undefined \def \showDOI       #1{#1}\fi
\ifx \showISBNx    \undefined \def \showISBNx     #1{\unskip}     \fi
\ifx \showISBNxiii \undefined \def \showISBNxiii  #1{\unskip}     \fi
\ifx \showISSN     \undefined \def \showISSN      #1{\unskip}     \fi
\ifx \showLCCN     \undefined \def \showLCCN      #1{\unskip}     \fi
\ifx \shownote     \undefined \def \shownote      #1{#1}          \fi
\ifx \showarticletitle \undefined \def \showarticletitle #1{#1}   \fi
\ifx \showURL      \undefined \def \showURL       {\relax}        \fi
\providecommand\bibfield[2]{#2}
\providecommand\bibinfo[2]{#2}
\providecommand\natexlab[1]{#1}
\providecommand\showeprint[2][]{arXiv:#2}

\bibitem[rep(tory)]%
        {repository}
 \bibinfo{year}{Repository}\natexlab{}.
\newblock
  \bibinfo{title}{\url{https://github.com/apnet23-reducing/evaluation}}.
\newblock
\newblock


\bibitem[Avin et~al\mbox{.}(2020)]%
        {sigmetrics20complexity}
\bibfield{author}{\bibinfo{person}{Chen Avin}, \bibinfo{person}{Manya Ghobadi},
  \bibinfo{person}{Chen Griner}, {and} \bibinfo{person}{Stefan Schmid}.}
  \bibinfo{year}{2020}\natexlab{}.
\newblock \showarticletitle{On the complexity of traffic traces and
  implications}.
\newblock \bibinfo{journal}{\emph{Proceedings of the ACM on Measurement and
  Analysis of Computing Systems}} \bibinfo{volume}{4}, \bibinfo{number}{1}
  (\bibinfo{year}{2020}), \bibinfo{pages}{1--29}.
\newblock


\bibitem[Benson et~al\mbox{.}(2010)]%
        {network_traffic_characteristics}
\bibfield{author}{\bibinfo{person}{Theophilus Benson}, \bibinfo{person}{Aditya
  Akella}, {and} \bibinfo{person}{David~A Maltz}.}
  \bibinfo{year}{2010}\natexlab{}.
\newblock \showarticletitle{Network traffic characteristics of data centers in
  the wild}. In \bibinfo{booktitle}{\emph{Proceedings of the 10th ACM SIGCOMM
  conference on Internet measurement}}. \bibinfo{pages}{267--280}.
\newblock


\bibitem[Chen et~al\mbox{.}(2013)]%
        {osa}
\bibfield{author}{\bibinfo{person}{Kai Chen}, \bibinfo{person}{Ankit Singla},
  \bibinfo{person}{Atul Singh}, \bibinfo{person}{Kishore Ramachandran},
  \bibinfo{person}{Lei Xu}, \bibinfo{person}{Yueping Zhang},
  \bibinfo{person}{Xitao Wen}, {and} \bibinfo{person}{Yan Chen}.}
  \bibinfo{year}{2013}\natexlab{}.
\newblock \showarticletitle{OSA: An optical switching architecture for data
  center networks with unprecedented flexibility}.
\newblock \bibinfo{journal}{\emph{IEEE/ACM Transactions on Networking}}
  \bibinfo{volume}{22}, \bibinfo{number}{2} (\bibinfo{year}{2013}),
  \bibinfo{pages}{498--511}.
\newblock


\bibitem[Li et~al\mbox{.}(2022)]%
        {patent}
\bibfield{author}{\bibinfo{person}{Weiqiang Li}, \bibinfo{person}{Rui Wang},
  {and} \bibinfo{person}{Jianan Zhang}.} \bibinfo{year}{2022}\natexlab{}.
\newblock \bibinfo{title}{Configuring data center network wiring}.
\newblock \bibinfo{howpublished}{US Patent 11,223,527}.
\newblock


\bibitem[Zhao et~al\mbox{.}(2019)]%
        {minimal_rewiring}
\bibfield{author}{\bibinfo{person}{Shizhen Zhao}, \bibinfo{person}{Rui Wang},
  \bibinfo{person}{Junlan Zhou}, \bibinfo{person}{Joon Ong},
  \bibinfo{person}{Jeffrey~C Mogul}, {and} \bibinfo{person}{Amin Vahdat}.}
  \bibinfo{year}{2019}\natexlab{}.
\newblock \showarticletitle{Minimal Rewiring: Efficient Live Expansion for Clos
  Data Center Networks.}. In \bibinfo{booktitle}{\emph{NSDI}}.
  \bibinfo{pages}{221--234}.
\newblock


\end{thebibliography}



\end{document}